# Models of Quiescent Black Hole and Neutron Star Soft X-Ray Transients

Insu Yi[1,2], Ramesh Narayan[1], Didier Barret[1], & Jeffrey E. McClintock[1]

[1] Harvard-Smithsonian Center for Astrophysics, 60 Garden St., Cambridge, MA 02138, USA
[2] Institute for Advanced Study, Olden Lane, Princeton, NJ 08540, USA

**Abstract.** When the mass accretion rate onto a black hole (BH) falls below a critical rate, $\dot{M}_{crit} \sim \alpha^2 \dot{M}_{Edd}$, accretion can occur via a hot optically thin flow where most of the dissipated energy is advected inward. We present such an advection-dominated model for the soft X-ray transient (SXT) A0620-00. This source has a puzzlingly low X-ray luminosity in quiescence, $\sim 6 \times 10^{30}$ erg s$^{-1}$, despite a relatively high mass accretion rate $\sim 10^{-10}$ $M_\odot$ yr$^{-1}$ deduced from its optical flux. The accreting gas makes a transition from a standard thin disc at large radii to an advection-dominated flow at small radii. The transition occurs when the effective temperature of the thin disc is $\sim 10^4$ K. Because of the very low accretion efficiency, $\sim 10^{-3} - 10^{-4}$, in the inner flow, the model fits both the optical and X-ray data. We also present models for V404 Cyg and Nova Mus 1991 in quiescence. Quiescent neutron star (NS) transients are expected to appear very different from BH systems because the advected energy is re-radiated from the NS surface whereas a BH swallows the advected energy. We discuss models for NS SXTs.

## 1. Introduction: Soft X-ray Transients in Quiescence

Soft X-ray transients (SXTs) are mass transfer binaries which undergo outbursts during which their luminosities increase by several orders of magnitudes. Currently, the quiescent state of SXTs lacks a satisfactory explanation. The problem of SXTs in quiescence has recently become more acute with the detection of X-ray emission in A0620-00 (McClintock et al. 1995). The optical luminosity of A0620-00 suggests a mass accretion rate $\dot{M} \sim 10^{-10} M_\odot$ yr$^{-1}$ in the outer parts of the accretion disc, while the observed X-ray luminosity implies

*Send offprint requests to*: I. Yi

$\dot{M} \sim 2 \times 10^{-15} M_\odot$ yr$^{-1}$ in the inner regions of the disc. However, at an accretion rate of $\dot{M} \sim 2 \times 10^{-15} M_\odot$ yr$^{-1}$, a standard thin accretion disc will be too cool to radiate at a temperature of 0.2 keV as observed in A0620-00. Thus, it is impossible to fit both the X-ray luminosity and the X-ray spectrum of A0620-00 with any standard thin disc model.

Assuming that the X-rays are produced close to the accreting BH and that the optical emission comes from farther out, the data suggest that the gas near the center must be very hot. Further, the X-ray luminosity is less than the optical luminosity, whereas in a normal accretion flow with a constant mass accretion rate we would expect the X-ray luminosity to be much greater since most of the gravitational potential energy of the accreting gas is released near the BH. Therefore, we deduce that either the mass accretion rate at the X-ray-emitting radii is much smaller than at the optical radii or that the radiative efficiency of the hot material is much less than that of the optical-emitting gas. We use the following scaled units for mass, mass accretion rate, and radius; $m = M/M_\odot$, $\dot{m} = \dot{M}/\dot{M}_{Edd} = \dot{M}/1.39 \times 10^{18} g s^{-1}$, and $r = R/R_S = R/2.95 \times 10^5 m$ cm.

## 2. A New Model: Cool Disc and Hot Advection-Dominated Flow

We propose a new model for the quiescent state of SXTs which provides a natural explanation of the optical and X-ray data (for details, see Narayan, McClintock, & Yi 1996). Our model is based on the idea that the accretion flow in quiescent SXTs has two distinct zones: (i) a standard thin disc at large radii which produces the optical and UV radiation, and (ii) a hot quasi-spherical flow in the inner regions which produces the X-rays. The hot inner flow corresponds to an optically thin solution of the accretion equations discovered recently (Narayan & Yi 1994, 1995a,b and references therein). This solution is viscously and thermally stable, in contrast to other hot accretion flow solutions discussed previously in the liter-ature. The newly discovered flow is *advection-dominated*, which means that most of the viscously dissipated energy is carried with the accreting gas as internal energy rather than being radiated.

We model the inner advection-dominated flows as follows. (1) We adopt the $\alpha$ viscosity prescription. (2) We assume quasi-equipartition magnetic fields in a two-temperature plasma. A constant fraction $(1-\beta)$ of the pressure is magnetic. $\beta = 0.95$ is assumed. Varying $\beta$ from 0.95 to 0.5 affects the results by only about 10%. (3) The self-consistently determined radiation efficiency $1-f$ is very low since the BH swallows most of the advected energy. The hot advection-dominated accretion flows are limited to accretion rates below the critical accretion rate, $\sim \alpha^2$ for BHs and $\sim 0.1\alpha^2$ for NSs (Narayan & Yi 1995b). (4) The cooling of the plasma via electrons lead to several emission components. The synchrotron emission gives a peak in optical range. The Comptonization of the synchrotron photons leads to a second peak in the soft X-rays. The combination of multiply scattered synchrotron photons and bremsstrahlung photons contribute to the extended hard X-ray peak. In the case of NSs, soft photons radiated from stellar surface Compton-cool hot electrons (Narayan & Yi 1995b). (5) For given $M$ and $\beta$, the inner flow has only one adjustable combination of parameters, $\dot{M}/\alpha$.

The flow at large radii corresponds to a standard thin accretion disc which is truncated at an inner radius $R_{in} \gg R_S$. Optical–UV emission is dominated by the outer thin disc. The luminosity $(=L_{out} \sim GM\dot{M}/R_{in})$ and optically thick emission spectrum are determined by $\dot{M}$, $R_{in}(=r_{in}R_S)$, and the outer disc radius $R_{out}(=r_{out}R_S)$. A constant $\dot{M}$ in the outer and inner regions is assumed. We determine $r_{out}$ or the projected Keplerian velocity $v_{out}$ following the procedure described by Smak (1981). We have used $r_{in}$ or the projected Keplerian velocity $v_{in}$ as a free parameter. The hot inner flow is advection-dominated and its radiative efficiency is only $\sim 10^{-4}-10^{-3}$, i.e. $L \sim (10^{-4}-10^{-3})GM\dot{M}/R_S$. This explains the low X-ray luminosity observed in these sources.

## 3. Quiescent Black Holes

### 3.1. A0620-00

We adopt the following parameters (Narayan et al. 1996). $M = 4.4\, M_\odot$, $v_{out} = 550 \pm 10$ km s$^{-1}$ or $r_{out} = 1.3 \times 10^5$, $r_{in} < 9000$, and $\alpha = 0.1$. We adjust $\dot{M}$ in the inner two-temperature flow so as to fit the X-ray flux. This gives $\dot{M} = 6.5 \times 10^{-5} \dot{M}_{Edd} = 6.3 \times 10^{-12} M_\odot$ yr$^{-1}$. We then use the same value of $\dot{M}$ for the outer thin disc and adjust the inner radius of this disc $r_{in}$ so as to fit the optical data. The fit gives $v_{in} = 5000$ km s$^{-1}$. The model agrees with all the measured fluxes, and is consistent with the upper limits. The model predicts a substantial flux in hard X-rays out to about 1 MeV (upper panel of Fig. 1).

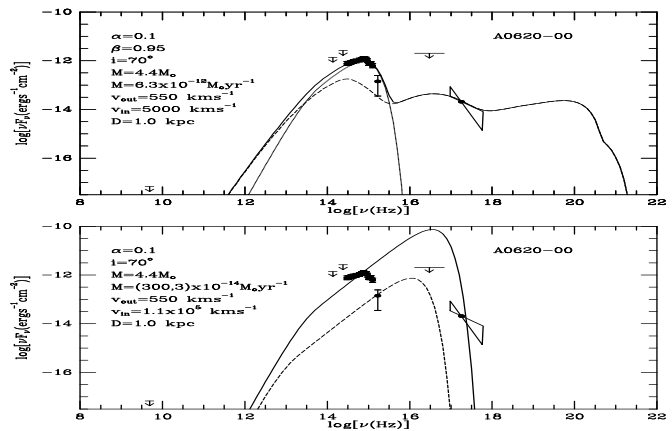

**Fig. 1.** Upper panel: Best fit to the composite spectrum of A0620-00. The filled circles are measured fluxes and arrows are upper limits. The ROSAT point at $\nu \sim 10^{17}$ Hz is shown with allowed spectral slopes indicated by the bow-tie. The adopted parameters are shown in the figure. The dotted line is the contribution from the outer disc and the dashed line is the contribution from the inner hot flow The total spectrum is the solid line. Lower panel: Possible fits in the standard thin disc model with no inner hot flow. The standard thin disc extends down to the marginally stable orbit. Two different mass accretion rates are adopted.

The models are able to fit the optical and X-ray data simultaneously. The spectrum of A0620-00 must turn up rather suddenly between UV and X-rays. Our model produces this upturn by having different components produce the optical and X-ray emission. While we have adjusted $\dot{M}/\alpha$ so as to fit the X-ray flux, the X-ray spectrum follows without further tuning. Although we treat $\alpha$ as a fitting parameter, we find that the best value we obtain is $\alpha \sim 0.1-0.3$, which is similar to estimates of $\alpha$ in hot discs in CVs during outburst. The value obtained for $\dot{M}$ is in the range $\sim 7 \times 10^{-12} - 2 \times 10^{-11} M_\odot$ yr$^{-1}$ which is within a factor of a few of the estimated mass storage rate (McClintock et al. 1983). The orbital velocity at the inner edge of the outer disc is in the range $\sim 3000-5000$ km s$^{-1}$, a range close to that inferred from the H$\alpha$ line profiles.

The model predicts a substantial luminosity in hard X-rays, an interesting test for future telescopes. As shown in the lower panel of Fig. 1, the standard thin disc models, where the thin disc extends down to the marginally stable orbit at $r_{in} = 3$, do not fit the data. The solid line shows a model where we adjusted $\dot{M}$ to pass through the optical and X-ray data. This model disagrees quite severely with the UV measurement and the EUV upper limit. On the other hand, if we reduce $\dot{M}$ so as to fit the UV point (dashed line), the fit in the optical and X-rays is poor.

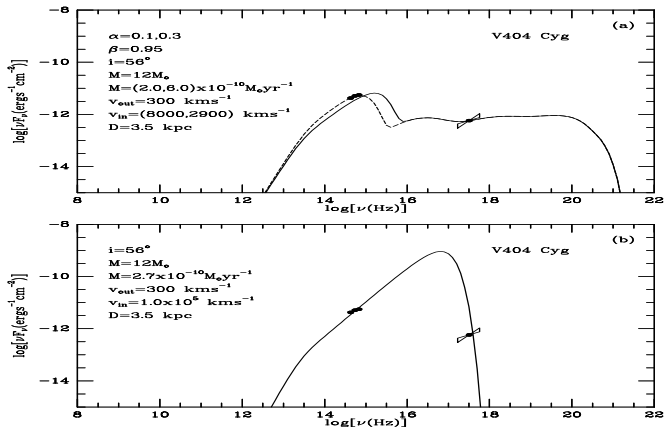

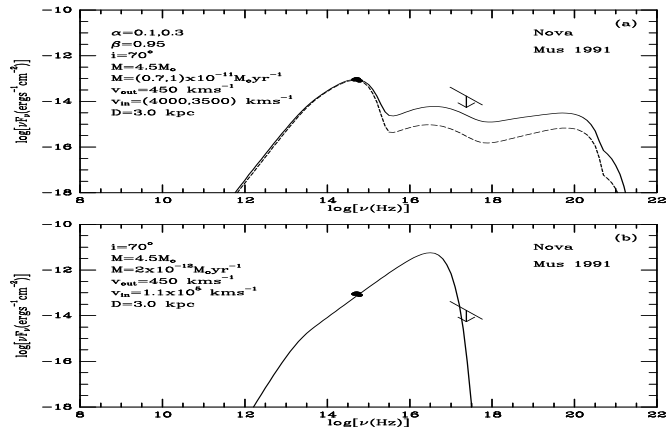

**Fig. 2.** (a) Two models of V404 Cyg corresponding to $\alpha = 0.1$ (solid) and $\alpha = 0.3$ (dashed). The adopted best fit model parameters are shown in the figure. (b) The best fit thin disc model where the disc is allowed to extend down to the marginally stable orbit. The two models can be easily distinguished by observations of $2-5$ keV X-rays or by EUV limits.

**Fig. 3.** (a) Two models of Nova Mus corresponding to $\alpha = 0.1$ (solid) and $\alpha = 0.3$ (dashed). The model parameters are shown in the figure. Both are consistent with available optical data and X-ray upper limit. (b) The best fit thin disc model where the disc extends down to the marginally stable orbit.

### 3.2. V404 Cyg

We adopt $M = 12\,M_\odot$, and $v_{in} \geq 1140$ km s$^{-1}$. Because of the absence of UV data for V404 Cyg, we are unable to determine $\dot{M}, \alpha$ and $v_{in}$ simultaneously from the data. Fig. 2(a) shows two models of V404 Cyg, where we have tried $\alpha = 0.1$ and $0.3$. In both cases, we are able to fit the optical and X-ray data reasonably well.

Due to the lack of UV data to constrain the size of the outer thin disc, in principle the data on V404 Cyg can also be fit with a standard thin disc model. Fig. 2(b) shows such a model, where we have adjusted $\dot{M}$ so as to fit the optical data points. The model gives an acceptable fit even to the X-ray flux. However, it makes a completely different prediction for the UV and EUV flux. Measurements in the UV are very difficult, given the large extinction ($A_v \sim 4$ mag) towards V404 Cyg. However, a limit on the EUV flux using the HeII 4686 line appears feasible and may provide a means of discriminating between the models. Most promising as a test to distinguish between the two classes of models is an X-ray observation with ASCA which could detect V404 Cyg to at least 5 keV if our model is correct.

### 3.3. X-Ray Nova Mus 1991

We adopt $M = 4.5\,M_\odot$, $v_{out} = 450 \pm 10$ km s$^{-1}$ and $v_{in} \geq 2000$ km s$^{-1}$. We have only an upper limit for the X-ray flux. In contrast to V404 Cyg, however, the optical data do provide both a flux and a nominal temperature for the thin disc. We use the optical data and estimate both $\dot{M}$ and $r_{in}$ simultaneously. Since the optical data cover only a small range of frequencies, we cannot measure the temperature very well and so the estimates of the individual values of $\dot{M}$ and $R_{in}$ are not very accurate. Nevertheless, we take our best-fit parameters and calculate the emission from the hot inner flow for $\alpha = 0.1$ and $0.3$. Fig. 3(a) shows the predicted X-ray flux is comfortably below the measured upper limits. Both models are consistent with the available data.

We also compute a standard thin disc model and the spectrum is shown in Fig. 3(b). This spectrum does not fit the shape of the optical data well. In addition, we see that it differs considerably from the two-zone models of Fig. 3(a) in the UV. Observations in the UV should be able to distinguish between the models.

### 3.4. Spectral Transition in Black Hole SXTs

When an SXT undergoes outburst, the luminosity increases by many orders of magnitude from quiescence to outburst, implying that the mass accretion rate onto the BH increases by several orders of magnitude from the quiescent mass accretion rate which we have estimated. As the mass accretion rate increases, the accretion flow may make a transition from a quiescent state, which we have described, to a thin disc flow extending down to the horizon. Since the hot advection-dominated accretion flow exists only up to the critical accretion rate ($\sim \alpha^2$), the transition will occur as the mass accretion rate increases. In this picture, we expect the spectrum to be hard in the early phase of outburst but to get softer near the peak of outburst when the entire accretion flow becomes a thin disc (Fig. 4). Similarly, during the decline from the peak of the outburst, we expect the soft spectral component to drop rapidly and the hard component to last longer.

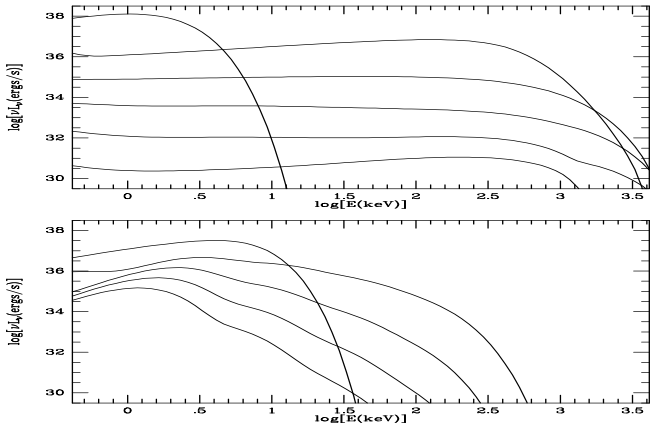

**Fig. 4.** Upper panel: Emission spectra from a $M = 10 M_\odot$ BH. $(\dot{m}, r_{in})$ varies from bottom to top as $(10^{-3.0}, 10^{3.2})$, $(10^{-2.5}, 10^{2.9})$, $(10^{-2.0}, 10^{2.6})$, $(10^{-1.5}, 10^{2.3})$, $(10^{-1.0}, 10^{2.0})$, $(10^{-0.5}, 10^{0.5})$. The top spectrum corresponds to a thin disk extending down to the marginally stable orbit. Lower panel: Emission spectra from a $M = 1.4 M_\odot$ NS. $(\dot{m}, r_{in})$ varies from bottom to top as $(10^{-3.0}, 10^{3.0})$, $(10^{-2.5}, 10^{2.5})$, $(10^{-2.0}, 10^{2.0})$, $(10^{-1.5}, 10^{1.5})$, $(10^{-1.0}, 10^{0.5})$. The top spectrum assumes only the thin disk. The spectral difference between the BHs and the NSs is mainly due to soft photons from NS surfaces which Compton cools hot advection-dominated flows. The NSs have softer spectra than those of BHs.

## 4. Advection-Dominated Flows in Quiescent Neutron Star Transients?

NS transients differ greatly from BH transients mainly due to the presence of the hard stellar surface in the former. In NS systems, the energy advected inward is eventually radiated from the surface, most likely in the form of soft photons. Therefore, for a given mass accretion rate, the NS's accretion efficiency is much higher than that of a BH when accretion occurs via an advection-dominated flow. The soft photons cool electrons via Comptonization and hence the emission temperature of electrons in the advection-dominated flow becomes lower than that in the BH systems (Narayan & Yi 1995b). The critical mass accretion rate ($\sim 0.1\alpha^2$) below which the advection-dominated flow exists is also substantially lower for NS systems primarily due to the efficient Compton cooling by soft photons. The resulting spectra from NS systems are much softer than those from BH systems (Fig. 4). If accretion occurs in the form of advection-dominated flow at low mass accretion rates, NSs are distinguishable from BHs by their spectral hardness.

## 5. Summary and Conclusions

(1) The emission from the inner hot flow has a number of different components. Most importantly, Compton upscattered synchrotron photons provide a significant luminosity in 0.1 − 1 keV X-rays, explaining the X-ray flux observed in A0620-00.

(2) Our results indicate that $\alpha$ lies in the range $\sim 0.1-0.3$. We deduce $\dot{M} \sim 10^{-11} M_\odot \text{yr}^{-1}$ in A0620-00 and Nova Mus, and $\dot{M} \sim$ few $\times 10^{-10} M_\odot \text{yr}^{-1}$ in V404 Cyg. The third parameter, $r_{in}$, lies in the range 3000-5000 $R_S$. The values of $r_{in}$ from our best-fit models all correspond to truncation at an effective surface temperature of 10000 − 15000 K. The idea of a thin disc evaporating to form a corona has been discussed recently by Meyer & Meyer-Hofmeister (1994) in the context of CVs.

(3) Our advection-dominated inner flow explains the low X-ray luminosities of quiescent BH SXTs. It is therefore important to allow for the low efficiency of advection-dominated flows when estimating the mass accretion rates of accreting BHs. The SXT models for BH transients as well as the model of Sgr A* described in Narayan, Yi, & Mahadevan (1995), *require a central horizon* through which all the advected energy disappears before it can be radiated.

(4) In the advection-dominated flows, the X-ray luminosity is not proportional to $\dot{M}$, but to $\dot{M}^2$. This is because the dominant cooling process is usually via Comptonization of synchrotron photons, which varies as the square, or a higher power, of the density. This effect provides an explanation for the large difference in luminosity between V404 Cyg and A0620-00 (or Nova Mus) despite their comparable $\dot{M}$'s.

(5) During the outburst, $\dot{M}$ onto the BH increases by several orders of magnitude from the quiescent values. Increased $\dot{M}$ may lead to observable spectral transitions. In the case of NS transients, soft radiation from the NS surface cools the advection-dominated flow, resulting in softer spectra.

*Acknowledgements.* RN was supported in part by NASA grant NAG 52837.